\documentclass[aps,prd,a4paper,twocolumn,showpacs,showkeys,preprintnumbers,amsmath,amssymb,nofootinbib]{revtex4}
\usepackage{graphicx}
\usepackage{bm}
\usepackage{dcolumn}
\usepackage{color}
\usepackage{pst-plot}

\usepackage{mathrsfs} 
\usepackage{txfonts}

\usepackage{footmisc} 

\newcommand{\be}{\begin{eqnarray}}
\newcommand{\ee}{\end{eqnarray}}

\def\beq{\begin{equation}}
\def\eeq{\end{equation}}
\def\ln{\,\mbox{log}\,}

\def\Res{\,\mbox{Res}\,}
\renewcommand{\Re}{\,\mbox{Re}\,}
\renewcommand{\Im}{\,\mbox{Im}\,}

\newcommand{\eq}[1]{(\ref{#1})}
\newcommand{\n}[1]{\label{#1}}

\def\al{\alpha}
\def\be{\beta}

\def\Ga{\Gamma}

\def\de{\delta}

\def\vp{\varepsilon}

\def\ka{\kappa}

\def\La{\Lambda}

\begin{document}

\title{Newtonian potential in higher-derivative quantum gravity} 

\author{Nicol\`o Burzill\`a}
\email{nburzilla@outlook.it}

\author{Breno L. Giacchini}
\email{breno@sustech.edu.cn}

\author{Tib\'{e}rio de Paula Netto}
\email{tiberio@sustech.edu.cn}

\author{Leonardo Modesto}
\email{lmodesto@sustech.edu.cn}

\affiliation{
{\small Department of Physics, Southern University of Science and Technology,}\\ 
{\small Shenzhen 518055, China}
}


\begin{abstract} \noindent
We hereby derive the Newtonian metric potentials for the fourth-derivative gravity including the one-loop logarithm quantum corrections. 
It is explicitly shown that the behavior of the modified Newtonian potential near the origin is improved respect to the classical one, but this is not enough to remove the curvature singularity in $r=0$. 
Our result is grounded on a rigorous proof based on numerical and analytic computations.

\end{abstract}


\maketitle
\noindent


\section{Introduction}
\label{Sec1}

Higher-derivative extensions of general relativity are ubiquitous when considering quantum gravity or quantum field theory in curved spacetimes \cite{UtDW}. While the quantum version of general relativity is perturbatively non-renormalizable, in~\cite{Stelle77} it was formally shown that the model with four derivatives of the metric is renormalizable. Regarding the semiclassical approach, the inclusion of higher-derivative terms in the gravitational sector is necessary to renormalize the vacuum diagrams, even though gravity is still treated as an external classical field (for a pedagogical review, see \cite{reviewS}). In this sense, the minimal higher-derivative gravity model needed to address the problem of renormalizability is the fourth-derivative gravity,
\beq
\n{HD}
S = S_{\text{EH}} +  \int d^4 x \sqrt{-g} \left\{ a_1 C^2 + a_2 E + a_3 \Box R + a_4 R^2
\right\}, 
\eeq
where  
\beq
S_{\text{EH}} = - \frac{1}{16 \pi G} \int d^4 x \sqrt{-g} \left(R+ 2\La \right) 
\eeq
denotes the Einstein-Hilbert action with cosmological constant, $C^2 = C_{\mu\nu\al\be} C^{\mu\nu\al\be}$ 
is the square of the Weyl tensor 
and $E$ 
is the integrand of the topological Gauss-Bonnet term in four dimensions.

The theory \eq{HD} shows good quantum properties like renormalizability and asymptotic freedom \cite{AvraBavi85}, but also ghost instabilities in its original quantization based on the Feynman prescription \cite{Stelle77}. However, a new quantum prescription \cite{AnselmiPiva1, AnselmiPiva2}, introduced recently by Anselmi and Piva, makes it possible to tame the ghost instability of the Stelle's model \eq{HD}. The prescription consists in a completion of the one proposed by Cutkosky, Landshoff, Olive and Polkinghorne (CLOP)
long ago \cite{CLOP} for the Lee-Wick theories \cite{LW1, LW2} and the unitary problem is now definitely solved at any perturbative order in the loop expansion \cite{AnselmiPiva3}. 
More precisely, the prescription involves deforming also the integration domain in the space of complex spatial momenta in order to make vanish the imaginary part of the amplitude at quantum as well as classical level. Furthermore, the procedure guarantees Lorentz invariance at quantum level, which was violated in the original incomplete CLOP prescription.

At the classical level, the ghost (or, in general, what the authors defined ``fakeons'') is removed solving the equations of motion for the fake fields by the mean of advanced plus retarded Green's function and fixing to zero the homogeneous solution \cite{ClassicalPrescription1, ClassicalPrescription2}. 
Such a prescription is very general and can be applied to ghosts or normal particles. In particular, it is needed to make perturbatively unitary 
the theory proposed by Modesto and Shapiro in \cite{ModestoShapiroLeeWick, ModestoLeeWick}, named ``Lee-Wick quantum gravity''.
Therefore, we can claim to have a class of finite and unitary theories of quantum gravity. The theory by Modesto and Shapiro has been designed to show up only complex conjugate poles (besides the graviton) in the propagator in order to guarantee tree-level unitarity. At the same time, the new prescription of \cite{AnselmiPiva3} ensures unitarity at any perturbative order.  The Wick rotation issue is also properly addressed by choosing the usual integration contour proposed by Lee and Wick \cite{LW1,LW2} and performing the integral on the energy $E$. This is
done by means of the residue theorem in the energy complex plane and the contributions on the arcs vanish as a particular feature of the dimensional regularization, as rigorously proved for any local higher-derivative theory in \cite{Aglietti:2016pwz}. 
In short, we can here summarize the proof given in the appendix A of \cite{Aglietti:2016pwz}. Let us consider the integral on the arc of the first quadrant of the energy complex plane. The latter is performed, as usual, fixing the radius $| E |$ and integrating on the angular variable. So far the integral is finite. Therefore, we expand the loop integrals in powers of the energy for $| E | \rightarrow +\infty$ to finally end up with a finite number of integrals that diverge. However, since such integrals are polynomial in the spatial momentum ${\bf k}$, in dimensional regularization they turn out to be zero in $D-1$ dimensions. Notice that this procedure works because there is a finite number of integrals.
Let us also mention that the peculiarity of Stelle's theory \cite{Stelle77} with the Anselmi-Piva prescription is that it is the only strictly renormalizable theory of gravity, versus the theories proposed in~\cite{ModestoShapiroLeeWick} which are super-renormalizable or finite~\cite{AsoreyLopezShapiro}. 

In this paper, we focus on the occurrence of curvature singularities in the fourth-derivative gravity~\cite{Stelle77} with one-loop quantum corrections \cite{Vil-unicEA,Gorbar:2002pw,BaVi85}. 
Indeed, being associated with small-distance and high-density configurations, quantum effects and related higher derivatives may have an important role in scenarios for which the classical gravitational solutions possess singularities, for example, inside black holes or in the primordial universe. Calculations in these regimes are complicated even at the classical level owed to the non-linearity of the theory and the presence of higher derivatives. Despite the difficulties, static spherically symmetric solutions have been obtained for the classical Stelle's gravity \cite{Stelle15PRL}, and it was shown that there exists a family of regular solutions, but it is not associated with a positive-definite energy density. The asymptotically flat solutions that couple to a physical source contain singularities; this can be seen already at the linear approximation. Indeed, in the linearised limit, the modified Newtonian potential is finite, but the curvature invariants are singular~\cite{Stelle78}.
 
The present work considerations are still restricted to the linear level, but we provide a step forward in the  approximation of quantum effects. Our treatment of the logarithm corrections is general, in the sense that they can be originated from the effective approach to quantum gravity, from the fourth-derivative gravity treated as a fundamental quantum theory, or from the integration of loops of mater fields in a given quantum field theory, etc.
The quantum effective action of our interest has the general structure
\begin{equation} 
\begin{split}
\label{act}
\Ga 
=  & - \frac{1}{\varkappa^2} \int d^4 x \sqrt{-g} \,
\Big\{ \, 2  R 
 +  \al_2 \, C^2 - \tfrac{\al_0}{3} \, R^2
\\
&
+  \be_2 \,  C_{\mu\nu\al\be} \, \ln \big( \Box/\mu_2^2  \big) \, C^{\mu\nu\al\be}
- \tfrac{\be_0}{3} \,  R \, \ln \big( \Box/\mu_0^2 \big)  \, R
\Big\} 
\,,
\end{split} 
\end{equation}
where $\varkappa^2 =  32 \pi G$, $\mu_s$ ($s=0,2$) are renormalization group invariant scales, and $\beta_s$ are related to the $\overline{\text{MS}}$-scheme beta functions. Notice that the constants $\beta_s$ depend on the particular quantum field content of the theory (see, {\it e.g.}, \cite{BD,book}). 
The classical action~\eqref{HD} is recovered from~\eqref{act} as the particular case $\beta_s = 0$. Here, we omitted the terms which are irrelevant for computing the Newtonian potential, such as the cosmological constant, $\Box R$ and the Gauss-Bonnet term $E$; we have also defined the coefficients in a more convenient manner for our purposes.

In the high-energy domain, where singularities may arise, the dominant finite quantum corrections to the effective action take the form of the logarithmic terms written in~\eqref{act}, since in this regime the quantum fields can be treated in a good approximation as massless. For instance, in \cite{Gorbar:2002pw} it was shown by the explicit calculation of the non-local $O(R^2)$ part of the effective action that, for a high-energy ($k^2 \gg m^2$) massive virtual quanta, the complicated one-loop form factors reduce to those in \eqref{act} (see also the discussion in \cite{Franchino-Vinas:2018gzr,Franchino-Vinas:2019upg}).
In the other extreme of the spectrum, the effective action also has the same general structure~\eqref{act}, although with different values for $\beta_s$. This happens because in the IR the massive degrees of freedom decouple according to the Appelquist and Carazzone theorem \cite{Appelquist:1974tg,Gorbar:2002pw}, while for massless fields we have an IR-UV correspondence of quantum effects. 

Another important
observation concerning the parameter $\be_s$ is 
that the contributions of different types of standard two-derivative matter fields have the same sign pattern. 
This also takes place in the fourth-derivative quantum gravity \cite{AvraBavi85} and in the effective quantum gravity based on the covariant definition of the effective action \cite{BaVi85} (see \cite{Giacchini:2020dhv} for recent developments and further references).
This universality of signs means that no cancellation of the beta functions is possible in the standard model or in beyond standard model theories which increases the number of scalar, vector and spinor fields, such as supersymmetry\footnote{ 
The same holds for the skew-symmetric rank-2 and rank-3 tensor fields, which may emerge after the compactification of the superstring effective action, 
since in this case, there is the quantum equivalence theorem with scalar and vector fields~\cite{Buchbinder:2008jf,Netto:2016imv}.
}.
A balanced phenomenological choice could be to couple Stelle's gravity, a renormalizable theory, to the standard model of particle physics. However, 
since we here want to keep the discussion on a general ground, we let the parameters $\beta_s$ arbitrary.

The paper is organized as follows. In section \ref{Sec2} we review the general perturbative formalism that relates the graviton perturbation to a static gravitational source (in particular, for the theory (\ref{act})) and the  
curvature invariants in terms of the gravitational perturbation. Moreover, we relate the perturbation $h_{\mu\nu}$ to two independent potentials. In section \ref{Sec4} we evaluate the potentials analytically near $r=0$, and exactly, but numerically, for the theory (\ref{act}). 
In section \ref{Sec6}, we compute the same potentials analytically but perturbatively at the first order in $\beta_s$. The latter computation considers only the first-order correction to the $2$-point  metric perturbation correlation function rather than the full re-summation of the one-loop $1$-particle irreducible dressed propagator like in (\ref{act}). Finally, in the last section we draw our conclusions.


\section{Singularities in the Newtonian limit}
\label{Sec2}
In the Newtonian limit we consider metric fluctuations around the Minkowski spacetime, \textit{i.e.},
\beq
\n{mli}
\sqrt{-g} g^{\mu\nu} \,=\, \eta^{\mu\nu} + \, \varkappa h^{\mu\nu},
\qquad |\varkappa h^{\mu\nu}| \ll 1 \, , 
\,
\eeq
and expand the action \eq{act} to the second order in $h_{\mu\nu}$.
We also introduce an action for the matter sector, whose energy-momentum tensor $T_{\mu\nu}$ couples linearly to gravity.
Therefore, the latter field sourced by $T_{\mu\nu}(x)$ can be evaluated utilizing the propagator, namely,
\begin{equation}
h_{\mu\nu}(x) = \frac{\varkappa}{2} \int d^4 x^\prime G_{\mu\nu \alpha\beta} (x - x^\prime) \, T^{\alpha\beta} (x^\prime).
\end{equation}
The Green's function $G_{\mu\nu \alpha\beta}$ is obtained by inverting the operator which follows from the expansion of the action up to second order in $h_{\mu\nu}$ subjected, of course, to a gauge-fixing condition. Finally, it is possible to show that the propagator associated to~\eqref{act} in the Euclidean signature reads (see, {\it e.g.},~\cite{Modesto12})
\beq
\label{prop}
G_{\mu\nu\al\be} (k)
=
 \frac{P^{(2)}_{\mu\nu\al\be}}{ k^2 f_2(k^2)}
-  \frac{P^{(0-s)}_{\mu\nu\al\be} }{ 2 k^2 f_0(k^2)} , 
\eeq
where the functions $f_s (z)$ for the quantum theory (\ref{act}) read
\beq \label{efizinho}
f_s (z) = 1 + z \, \big[  \al_s + \be_s \ln \big( z/\mu_s^2\big) \big]
\,,
\qquad s = 0, 2  \, . 
\eeq
Notice that in \eqref{prop} we omitted the gauge-dependent terms which are not relevant for the Newtonian limit.
Moreover, $k^2 = k_\mu k^\mu$ and $P^{(2)}$
and $P^{(0-s)}$
are the spin-$2$ and spin-$0$ projectors~\cite{Barnes-Rivers},
\begin{eqnarray}
P^{(2)}_{\mu\nu\alpha\beta}
& =&  \dfrac{1}{2} \big( \theta_{\mu\alpha}\theta_{\nu\beta}
+ \theta_{\mu\beta}\theta_{\nu\alpha} \big)
- \dfrac{1}{3} \theta_{\mu\nu}\theta_{\alpha\beta} \, ,
\\
P^{(0-s)}_{\mu\nu\alpha\beta}
 &=& 
\dfrac{1}{3} \theta_{\mu\nu}\theta_{\alpha\beta} \, ,
\end{eqnarray}
where the longitudinal and transverse vector-space projectors are respectively:
\beq
\omega_{\mu\nu} = \frac{k_\mu k_\nu}{k^2}
\qquad \text{and} \qquad
\theta_{\mu\nu} = \eta_{\mu\nu} - \frac{k_\mu k_\nu}{k^2}.
\eeq
Therefore, the solution associated with a point-like massive source at rest with energy-momentum tensor 
\beq
T_{\mu\nu}(\vec{r}) =  M \,  \de_\mu^0 \de_\nu^0 \, \de^{(3)} (\vec{r})
\eeq
is given by the diagonal metric perturbation with components:
\begin{eqnarray}
&& h_{00} = \frac{\varkappa M}{4} \left( \frac43 I_2 - \frac13 I_0 \right) , \\
&& h_{11} = h_{22} = h_{33} = \frac{\varkappa M}{4} \left( \frac23 I_2 + \frac{1}{3} I_0 \right) \, , \\ 
&& 
\n{IS}
I_s = \int \frac{d^3 k}{(2\pi)^3} \frac{e^{-i \vec{k}\cdot\vec{r}}}{\vec{k}^2 f_s(\vec{k}^2)} = - \frac{1}{2\pi^2 r} \int_0^\infty dk \frac{\sin(kr)}{k f_s(k^2)} \, , 
\end{eqnarray}
with $| \vec{k} | = k$.

By defining the auxiliary spin-$s$ potentials~\cite{BreTib1}
\beq
\n{chi-novo}
\chi_s(r) = \ka_s M I_s(r) , \quad \text{with} \quad \ka_s \equiv \frac{\varkappa^2}{8} \left( \frac{3s}{2} -1 \right) ,
\eeq
one can write the associated Newtonian potentials $\Phi$ and $\Psi$ in the form
\begin{align} 
\label{Phi}
& \Phi(r)  =  \frac{\varkappa }{2}\, h_{00}= \frac{1}{3} (2 \chi_2 + \chi_0),
\\
\label{Psi}
& \Psi(r) =  \frac{\varkappa }{2}\, h_{11}  = \frac{1}{3} (\chi_2 - \chi_0).
\end{align}
The main benefit of using the potentials~\eq{chi-novo} is because they only depend on the spin-$s$ sector of the propagator (see eqs.~\eqref{prop} and \eq{IS}) and allow to split the contributions owed to the scalar and tensor degrees of freedom. Since the Newtonian potentials $\Phi$ and $\Psi$ are linear combinations of $\chi_s$, there is no loss of generality in restricting our considerations to the latter ones. For example, if both $\chi_{0,2}$ are bounded, the metric is bounded too. Moreover, the structure of the equations~\eq{IS} and~\eq{chi-novo}, which define $\chi_{0,2}$, is essentially the same, allowing the derivation of general results by working with only one generic function $f_s$.

The occurrence of spacetime singularities in the solution~\eqref{Phi}--\eqref{Psi} can be investigated by checking whether the curvature invariants built with this linearised metric are bounded. For instance, for the Kretschmann scalar one gets:
\begin{equation} \label{Krets}
R_{\mu\nu\al\be}^2 = 
4 \big( \Phi^{\prime\prime 2} + 2\Psi^{\prime\prime 2} \big) 
+\frac{16}{r} \, \Psi^\prime \Psi^{\prime\prime }
+\frac{8}{r^2}\, \big( \Phi^{\prime 2} +3 \Psi^{\prime 2} \big) 
\,,
\end{equation}
while the Weyl tensor squared and the scalar curvature read
\beq \label{Weyl-R}
C_{\mu\nu\al\be}^2 = \frac{4}{3} \left( \chi_2^{\prime\prime } - \frac{\chi_2^{\prime}}{r} \right)^2
\quad \text{and} \quad
R = 2 \left( \chi_0^{\prime\prime } + \frac{2 \chi_0^{\prime}}{r} \right)
\,,
\eeq
which depend only on the spin-$2$ and spin-$0$ sectors, respectively.

In view of these equations, the existence of the limits
\beq \label{RegCond}
\lim_{r \rightarrow 0} \chi_s^{\prime\prime}(r) < \infty
\qquad \mbox{and} \qquad
\lim_{r \rightarrow 0} \frac{\chi_s^\prime(r)}{r} < \infty \, , 
\eeq
is a necessary and sufficient condition for avoiding curvature singularities\footnote{It is possible to show that these conditions ensure the regularity of the other invariants built with curvature tensors only, such as $R_{\mu\nu}^2$ (see, {\it e.g.},~\cite{BreTib1,Giacchini:2018zup}).}.
Therefore, the finiteness of the potentials is not enough to avoid curvature singularities. Let us point out that if the potential $\chi_s(r)$ is analytic around $r=0$, then the first limit in~\eqref{RegCond} is automatically satisfied, whereas the condition posed by the second one reduces to $\chi_s^\prime(0) = 0$ (being an analytic function, this means that $\chi_s^\prime(r) \sim r$ for small enough~$r$). However, when dealing with non-analytic form factors of the type~\eqref{efizinho}, one must be careful and verify if the two conditions in~\eqref{RegCond} holds separately. Extending the terminology of~\cite{Frolov:Poly}, we shall say that a potential $\chi_s (r)$ is \textit{regular} if it regularises the curvature invariants, that is, if it satisfies~\eqref{RegCond}.
According to~\eqref{Weyl-R}, the potentials $\chi_0$ and $\chi_2$ are responsible, respectively, for the regularity of the scalars $R$ and $C_{\mu\nu\al\be}^2$, while to ensure that the Kretschmann scalar~\eqref{Krets} does not diverge, both potentials must be regular.

Before considering the effects of the logarithmic quantum corrections to the potential, it is instructive to remind the  classical fourth-derivative gravity results, \textit{i.e.}, for $\be_s = 0$ in~\eqref{act}. This theory is a well-known example of the existence of curvature singularities despite the potentials being finite. In fact, the integral in~\eqref{IS} in this case gives
\begin{equation} \label{Int4Der}
\int_{0}^{\infty} dk \, \frac{\sin(kr)}{k(1+\al_s k^2)}   =  \frac{\pi}{2} \left( 1 - e^{-m_s r} \right), 
\end{equation} 
where $m_s = \al_s^{-1/2}$ (hereafter, we assume $\al_s >0$, otherwise the particle with mass $m_s$ would be a tachyon). Whence,
\beq 
\label{Pot4d}
\chi_s(r) = \, -\frac{\ka_s M}{4\pi r} \left( 1 - e^{-m_s r} \right) . 
\eeq
The potential \eq{Pot4d} is finite~\cite{Stelle77}. Indeed,
\begin{equation}
\n{Pot4d2}
\chi_s(r) =  -\frac{\ka_s M}{4\pi} \left( m_s - \frac{m_s^2}{2} \, r  \right)  + O(r^2),
\end{equation}
but the second condition in~\eqref{RegCond} is violated because \eqref{Pot4d2} implies $\chi_s^\prime(0) \propto m_s^2$. Thus the potential~\eqref{Pot4d} is not regular, and the curvature invariants diverge as $r \to 0$~\cite{Stelle78}.

We recall that for local higher-derivative gravitational theories with \emph{more} than four derivatives in the spin-$s$ sector of the classical action, the associated potential $\chi_s(r)$ is not only finite \cite{Newton-MNS,Newton-BLG}, but also regular \cite{BreTib1}. This result was extended to a larger class of non-local ghost-free gravity models defined by analytic form factors in~\cite{BreTib2}. Therefore, all super-renormalizable models of refs.~\cite{AsoreyLopezShapiro,Tomboulis,Modesto12} have a regular Newtonian limit. The question of whether the insertion of the non-analytic logarithmic quantum correction spoils the good regularity features of these models is addressed in a separate publication~\cite{Nos6der}.
Given the difference between classical fourth- and higher-derivative gravity models in what concerns the presence of singularities, in the next section we will investigate if the improvement due to the logarithmic quantum correction is strong enough to regularize the curvature singularities that are present at the classical level.

To simplify the notation, in the following we will only write the $s$-label when this specification is necessary to avoid ambiguity.


\section{Towards Newtonian potentials in fourth-derivative quantum gravity}
\label{Sec4}
In order to get the Newtonian potentials for the theory~\eqref{act}, namely, the one-loop $1$-particle irreducible quantum corrections to $h_{\mu\nu}$, we have to evaluate 
the following integral, 
\beq \label{Chi-Int}
\chi (r) = - \frac{\ka M}{ 2\pi^2 r} \int_0^\infty dk
\frac{\sin (kr)}{k[1+\al k^2 + 2\beta k^2 \ln (k/\mu) ]} \, . 
\eeq
A natural attempt is to apply the methodology based on Cauchy's residue theorem, which was proved to be successful for theories whose propagator has massive poles~\cite{Stelle77,Newton-MNS,Newton-BLG,BreTib1}. 
The first step would be to identify the propagator's poles and define a contour $C$ in the complex plane such that the integral~\eqref{Chi-Int} can be obtained as part of the contour. Via the residue theorem, the value of the integral would then be related to the pole(s) inside $C$.

The massive poles of the propagator~\eqref{prop} are the zeros of the equation
\beq \label{EqZeros}
1  + k^2 \left[  \al + 2 \, \be \ln ( k / \mu ) \right] = 0 \, , 
\eeq
that has an infinite number of solutions because the complex logarithm is a multivalued function. 
A detailed study of the structure of these poles in the Riemann surface has been carried out in the ref.~\cite{Calmet:2017omb}.
However, we can focus on the principal branch because our final goal is to solve an integral of the type \eqref{IS} over the real line. Indeed, only in the principal branch the complex logarithm restricted to the real line coincides with the real one.
In this case, the eq.~\eqref{EqZeros} can have either one real root, or two complex conjugate roots~\cite{Calmet:2017omb,Calmet:2014gya,Calmet:2015pea,Calmet:2017rxl,Calmet:2018uub}, namely, $k^2 = -m^2$ and $k^2 = -\overline{m}^2$, where
\beq \label{Lambert}
m^2 = \frac{1}{\be \, W\left(- \frac{e^{\al/\be}}{\be\mu^2}\right)}
\eeq 
and $W$ is the Lambert (product logarithm) function. The quantity $m^2$ is complex provided that
\beq \label{ComplexCond}
e^{\frac{\alpha}{\beta} + 1} > \be \mu^2 \quad \text{with} \quad \be > 0.
\eeq
Notice that $m^2$ becomes real and negative (tachyonic) for $\be\mu^2 > e^{\frac{\alpha}{\beta}+1}$; it is also negative if $\be < 0 $, as $W(x)>0$ for $x>0$. In what follows we assume that the condition~\eqref{ComplexCond} is satisfied so as to avoid tachyon instabilities.

Nevertheless, it seems that the procedure based on the residue theorem does not help in the case of~\eqref{Chi-Int}. Choosing the logarithm's branch cut along the negative imaginary axis, it is possible to define a contour $C$ formed by a path that goes along the real axis (with an indentation around the origin) and which is closed by a semi-circular arc in the upper half-plane. However, the integral along the negative part of the real axis yields a term $\ln\vert z \vert+ i\pi$ in the integrand's denominator, which results in a new integral which seems to be more complicated than the original one. Other branch cuts and/or integration contours can be used, but it always remains an integral to be evaluated. The conclusion is that this procedure is not as useful as in the cases mentioned above.


\subsection{Small-$r$ behavior of the potential}
\label{Sec4.1}

Although we could not find an explicit expression for the potential $\chi(r)$, we can use the integral representation~\eqref{Chi-Int} to deduce its behavior near $r=0$ and study its regularity properties. For this purpose, let us change variables $k \mapsto 1/u$ in the integral in eq. \eq{IS},
\begin{equation} \label{26}
\int_{0}^{\infty} dk\, \frac{\sin(kr)}{kf(k^2)}  = \int_{0}^{\infty} du\, \frac{\sin(r/u)}{u f(1/u^2)} \, .
\end{equation}
Using the Schwinger parametrization\footnote{
That is, given an $x>0$, it holds
\begin{equation*}
\frac{1}{x} = \int_0^\infty d\xi\, e^{-x \xi } \, .
\end{equation*}
Since the Schwinger parametrization \eqref{I-defin} is going to be applied in the integral \eqref{26}, we must make sure that the effect of varying $u$ does not spoil the validity of the parametrization---which is equivalent to show that  $u f(1/u^2) > 0$ for $u>0$. This is indeed the case as we already assumed that the propagator has no tachyonic poles; therefore, being a continuous function, $f(1/u^2)$ does not change sign along the real line.}
it follows
\begin{equation}
\begin{split} 
\n{I-defin}
\frac{1}{u f(1/u^2)} & =  \frac{u}{u^2 + \al - 2 \be \ln \left( \mu \, u  \right) } 
\\
& = u \int_0^\infty d\xi \, (\mu \, u)^{2\be\xi}  e^{ -( \al + u^2) \xi  } \,  . 
\end{split} 
\end{equation}
Whence \eq{IS} turns into:
\begin{equation}
\begin{split} 
I = -\frac{1}{2\pi^2 r}\int_0^\infty  \int_0^\infty d \xi \, du \, u \sin\left( r/u \right)  (\mu u)^{2\be\xi}  e^{ -( \al + u^2) \xi  } 
.
\end{split} 
\end{equation}
Performing the integration in the variable $u$ and inserting the result into~\eqref{Chi-Int}, the potential reads
\begin{equation} \label{Chi-IntQ}
\chi(r) = - \frac{\ka M}{4 \pi^2} \int_0^\infty d\xi\, e^{-\al\xi} \left[ t_{1}(r,\xi) + t_{2}(r,\xi) \right] \, , 
\end{equation}
where
\begin{align}
& t_1(r,\xi) =   2 \, r (\mu r)^{2\be\xi} \, \Ga(-2-2\be\xi) \, \sin(\pi \be\xi)
\nonumber \\
& \hspace{1.3cm} 
\times \, {_0F_2}\left( \tfrac{3}{2} + \be\xi , 2 + \be\xi ; \tfrac{r^2 \xi}{4} \right) \, ,  \label{t1} \\
&
\label{t2}
t_2(r,\xi) =  \xi^{- \tfrac{1}{2}} \left( \tfrac{\mu^2}{\xi} \right)^{\be\xi}   \Ga( \tfrac{1}{2} + \be\xi) \, {_0F_2}\left(\tfrac{3}{2} , \tfrac{1}{2} - \be\xi ; \tfrac{r^2 \xi}{4} \right) \, . 
\end{align}
Moreover, ${_0F_2}\left( y_1 , y_2 ; z \right) = {_0F_2}\left( -; y_1 , y_2 ; z \right)$ is the generalized hypergeometric function.

Before discussing the small-$r$ behavior of this potential, it is useful to make another brief digression on the classical fourth-derivative gravity. In fact, the potential~\eqref{Pot4d} can be recovered from the three previous equations simply by taking the limit $\beta \to  0$. In this case 
\begin{equation}
t_1(r,\xi) =  -\frac{\pi r}{2}  \, {_0F_2}\left( \tfrac{3}{2}, 2; \tfrac{r^2 \xi}{4} \right) 
\end{equation}
gives the odd-power terms of the series expansion of $\chi(r)$, while 
\begin{equation} \label{t2-4der}
t_2(r,\xi) = \sqrt{\frac{\pi}{\xi}}  \,\, {_0F_2}\left(\tfrac{3}{2} , \tfrac{1}{2} ; \tfrac{r^2 \xi}{4} \right)
\end{equation}
gives the even-power ones. It is possible to integrate the series in $r$ term by term because ${_0F_2}$ is an entire function of $\xi$ and the exponential damping in~\eqref{Chi-IntQ} makes each term well-behaved in the limit of large $\xi$. The only divergent integrand in the series is the zero-order term, which behaves like  $\xi^{-1/2}$ for small $\xi$, see~\eqref{t2-4der}. However, the latter improper integral can be performed to finally get the value of $\chi(0)$ in eq.~\eqref{Pot4d}.

Following the intuition from the classical fourth-derivative gravity case, one may be tempted to analyse the terms related to $t_1$ and $t_2$ separately also in the more general case of eqs.~\eqref{t1} and~\eqref{t2}. It turns out that this procedure does not work because $t_1$ cannot be written as a standard power series in $r$ due to the term $r^{2\xi\be}$. Moreover, regarded as functions of $\xi$, $t_1$ and $t_2$ have discontinuities for finite values of $\xi$. In the case of $t_1$ this is generated by the gamma function with negative arguments, meanwhile in $t_2$ it is due to the occurrence of negative parameters in the hypergeometrical function. Indeed,
using the identity
\beq
\Gamma(x)\Gamma(-x)= - \frac{\pi}{x \sin(\pi x)}
\eeq
one obtains
\begin{equation} \nonumber
 \Ga(-2-2\be\xi)  \sin(\pi \be\xi) = - \frac{\pi}{4(1+\be\xi) \Ga(2+2\be\xi) \cos(\pi \be\xi)},
\end{equation}
implying that $t_1(r,\xi)$ only diverges at $\xi=\frac{1+2n}{2\beta}$ (with $n\in\mathbb{N}$). This is the same condition that defines the singular points in $t_2(r,\xi)$. 
Actually, only the sum of the two functions makes sense and gives a well-behaved integrand in eq.~\eqref{Chi-IntQ}. 
These features are related to the parameter $\be$ and, therefore, they are just a manifestation of the non-analyticity of the logarithmic function in~\eqref{Chi-Int}, which makes it not possible to write a power series of the resultant potential $\chi(r)$ around $r=0$. 

Despite the impossibility of finding a series representation to \eqref{Chi-IntQ}, we can still analyze its behavior for small $r$ and investigate the regularity of the Newtonian at the origin. For this purpose, let us divide the sum $t_1 + t_2$ in \eqref{Chi-IntQ} in a part dependent on $r$ and a part that does not depend. 
In order to achieve this goal, we define the following two functions,
\begin{eqnarray} 
&& t_0(\xi) \equiv t_2(0,\xi) =   \xi^{- \tfrac{1}{2}} \left( \tfrac{\mu^2}{\xi} \right)^{\be\xi} \, \Ga( \be\xi + \tfrac{1}{2} ) ,
\label{t0}
\\
&& \label{t}
t(r,\xi) =   - t_0(\xi) + t_1(r,\xi) + t_2(r,\xi) .
\end{eqnarray}
The function $e^{-\al\xi}t(r,\xi)$ of the variable $\xi$ is bounded because the singularities that $t_1$ and $t_2$ have for finite values of $\xi>0$ cancel each other, and the singularity that $t_2$ has for $\xi \to 0$ is cancelled by $t_0$. Furthermore, it is clear that $e^{-\al\xi}t(r,\xi)$ is small for $r^2\ll \min \lbrace \beta, \mu^{-2} \rbrace$, except for a region near $\xi = 0$. Indeed, for $\beta \xi >1/2$ the function $t(r,\xi)$ is at least of order $\mu r^2$ or $\beta^{-1/2}r^2$ and in a region near $r=0$ the leading contributions come from small $\xi$ too. Taking into account these two approximations, it follows:\footnote{Notice that we cannot expand  $(\mu r)^{2\be\xi} = 1 + 2\xi \beta \ln (\mu r) + 2 \xi^2 \beta^2  [\ln (\mu r)]^2 + O(\xi^3\beta^3)$ because each of these terms diverges when $r \to 0$; this gives an indeterminacy when $\xi$ is integrated down to 0. Ultimately, the behavior of the potential for small $r$ results from this ambiguity.}
\begin{equation} 
t(r,\xi) = -\frac{\pi r}{2} \,  (\mu r)^{2\be\xi} \,[  1  + c_1 \beta \xi + O(\xi^2) ] + O(r^2) \, ,
\end{equation}
where $c_1$ is a constant.

Comparing~\eqref{Chi-IntQ} with~\eqref{t}, the potential reads 
\begin{equation}
\chi(r) = -\frac{\ka M}{4 \pi^2} \int_0^\infty d\xi \, e^{-\al \xi} [t_{0}(\xi) + t(r,\xi)].
\end{equation}
Since it holds (assuming $\beta \ln(\mu r) < 0$,  $\alpha>0$ and $n \in \mathbb{N}$)
\begin{equation} 
\int_0^\infty d \xi \, e^{-\al \xi} (\mu r)^{2\be\xi}  \xi^n   = \frac{  n! }{[\alpha - 2\beta \ln (\mu r) ]^{n+1}} \, ,
\end{equation}
it is straightforward to verify that
\begin{equation} \label{ChiS_app}
\chi(r) = -\frac{\ka M}{8 \pi} \Bigg[ c_{0}  -  \frac{ r }{\alpha - 2\beta \ln ( \mu r ) } + O\left( r[\ln(\mu r)]^{-2}\right)  \Bigg] \, ,
\end{equation}
where 
\begin{equation}
c_{0} = \frac{2}{\pi} \int_0^\infty d \xi \, e^{-\al \xi} t_{0}(\xi)  .
\end{equation}
The potential is finite at $r=0$,
\begin{equation}
\lim_{r\to 0} \, \chi(r) = -\frac{\ka M c_{0}}{8 \pi}  ,
\end{equation}
nonetheless, as  discussed in Sec.~\ref{Sec2}, the regularity of the Newtonian limit is related to the behavior of the derivatives of $\chi (r)$ near the origin. Differentiating~\eqref{ChiS_app} it follows that 
\begin{equation} \label{DerChi_app}
\chi^\prime(r) \underset{r \to 0}{\sim } 
\frac{\ka M}{8 \pi} \, 
\frac{1}{\alpha - 2\beta \ln (\mu r)}.
\end{equation}
This expression coincides with the one presented in~\cite{FroVilkMG}, and shows that the potential satisfies $\chi^\prime(0) = 0$. However, it tends to zero in such a slowly manner that the curvature singularities still remain. In fact,
\begin{equation}
 \lim_{r\rightarrow 0} \frac{\chi^\prime(r)}{r} = \infty \, .
\end{equation}
It turns out that the solution~\eqref{ChiS_app} violates both the regularity conditions in~\eqref{RegCond}
because of its non-analyticity. Indeed, we also have:
\begin{equation} 
\chi^{\prime\prime}(r) \underset{r \to 0}{\sim } 
\frac{\ka M}{8 \pi} \, 
\frac{2\be}{r[\alpha - 2\beta \ln (\mu r)]^2} \, , 
\end{equation}
which diverges in the limit $r \to 0$.

This analysis reveals that the one-loop non-local quantum corrections to the fourth-derivative gravity do not substantially modify the regularity of the Newtonian potential. Unlike in the case of analytic form factors, the improvement $\chi^\prime(0) = 0$ is not enough to regularize the curvature invariants.


\subsection{Numerical analysis of the potential}

\begin{figure}
\begin{center}
\includegraphics[width=6.0cm,angle=0]{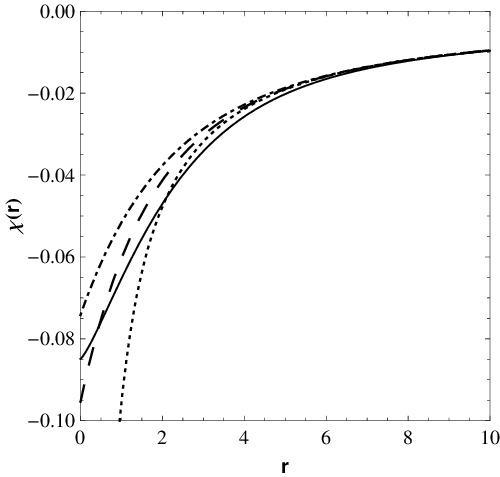}
\caption{ \sl Plot of $\chi(r)$ for different models: numerical solution of eq.~\eqref{Chi-Int} for $\al=1.0$, $\mu=1.2$ and $\be=0.5$ (solid line), classical  fourth-derivative gravity with $\al=1.0$ ($\be=0$) (dashed), the modified potential of eq.~\eq{C} (dot-dashed), and the $1/r$ Newton's potential ($\al=\be=0$) (dotted).}
\label{f1}
\vspace{3mm}
\includegraphics[width=6.0cm,angle=0]{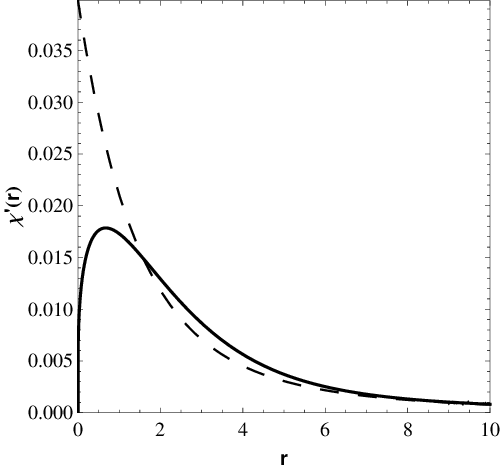}
\caption{\sl Derivative of the numerical solution (solid line) and of the classical fourth-derivative gravity (dashed). It is clear that the second derivative of the potential changes sign only for the log-corrected solution, which also satisfies $\chi^\prime(0)=0$.}
\label{f2}
\includegraphics[width=6.0cm,angle=0]{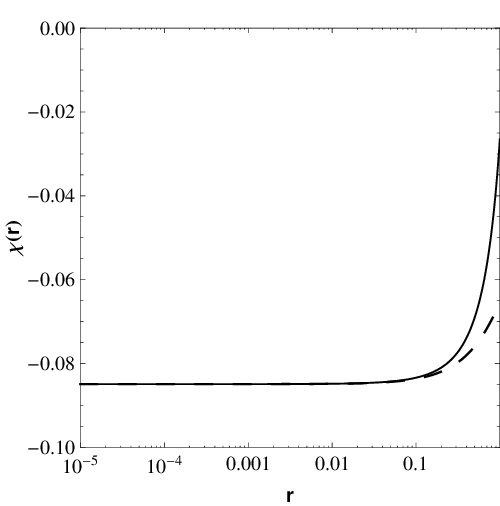}
\caption{\sl Comparison of the approximate solution~\eqref{ChiS_app} close to $r=0$ (solid line) with the numerical one (dashed) in the linear-log scale.}
\label{f3}
\end{center}
\end{figure}

Given the difficulties of finding an explicit expression for the fourth-derivative gravity potential with one-loop logarithmic quantum corrections, here we carry out a numerical analysis that serves as a double check of the general results proved in the previous section. In Fig.~\ref{f1} we plot the numerical integration of the potential~\eqref{Chi-Int}. 
In order to make a comparison of different theories, we display the Newton potential in Einstein's gravity, 
the solution~\eqref{Pot4d} for the classical fourth-derivative gravity, 
and the modified potential of ref.~\cite{Calmet:2018hfbKuipers:2019qby}. 
The latter, in our notations, has the form
\beq
\n{C}
\chi(r) =  \, -\frac{\ka M}{4\pi r} \left( 1 - e^{-\text{Re}(m) r} \right)
,
\eeq
where the effective mass $m$ is given by eq.~\eqref{Lambert}, see~\cite{Calmet:2018hfbKuipers:2019qby} for a further discussion.

The direct inspection of Fig.~\ref{f1} suggests that the corrected potential changes its concavity as it approaches $r=0$. This is clear in the plot of the derivative of the solutions in Fig.~\ref{f2}, which even shows that $\chi^\prime(0) = 0$, as we proved above in a general and analytic setting. 
These distinguishing features are not present in the potential of the classical fourth-derivative gravity, nor in its modification proposed in~\cite{Calmet:2018hfbKuipers:2019qby}. In fact, the modification \eq{C} does not change the functional form of the potential \eq{Pot4d}, but only its massive parameter.

The change of concavity of $\chi(r)$ can be verified analytically using the results obtained in the previous section.
Indeed, 
$\chi^\prime(r)$ vanishes for $r \to 0$ and $r \to \infty$, then, because of Rolle's theorem, there must be some $r_0 \in (0,\infty)$ such that $\chi^{\prime\prime}(r_0) = 0$. Obviously, we cannot investigate the potential for large $r$ making use of \eqref{ChiS_app}, but it should be analyzed by taking the limit of large $u$ in eq.~\eqref{I-defin}. This is equivalent of making a perturbative expansion in $\beta$, but we postpone this discussion to Sec.~\ref{Sec6}. We hereby only point out that in the limit $r \to \infty$ the integral~\eqref{I-defin} tends to $\pi/2$, giving the standard Newton's potential proportional to $r^{-1}$, as expected and suggested by Fig.~\ref{f1}.

Finally, in Fig.~\ref{f3} we compare the analytic small-$r$ approximation obtained in the previous section, given by eq.~\eqref{ChiS_app}, with the numerical solution of eq.~\eq{Chi-Int}. As expected, they agree with good precision near $r=0$.

\section{Perturbative solution of the potential and infrared limit}
\label{Sec6}

\begin{figure} 
\begin{center}
\includegraphics[width=7cm,angle=0]{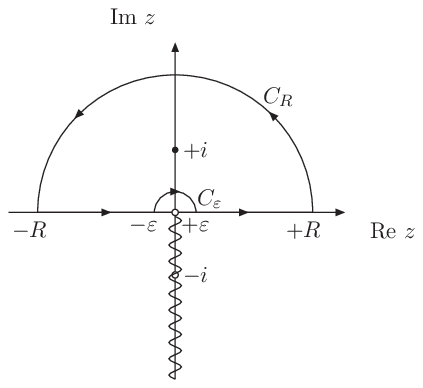}
\caption{\sl Contour of integration used to evaluate~\eqref{eq4}, poles and branch cut defined by~\eq{eqt1}.}
\label{con}
\end{center}
\end{figure}

\begin{figure} 
\begin{center}
\includegraphics[width=4.1cm,angle=0]{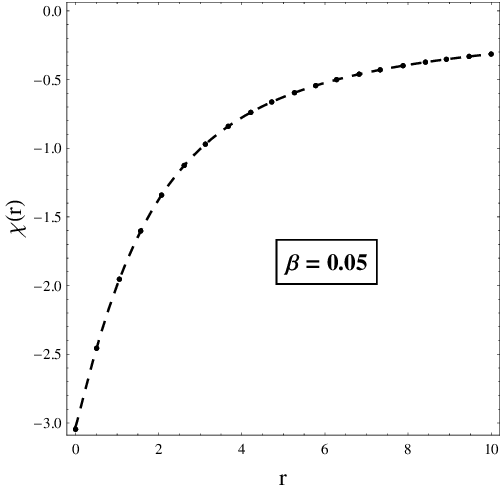}
\includegraphics[width=4.1cm,angle=0]{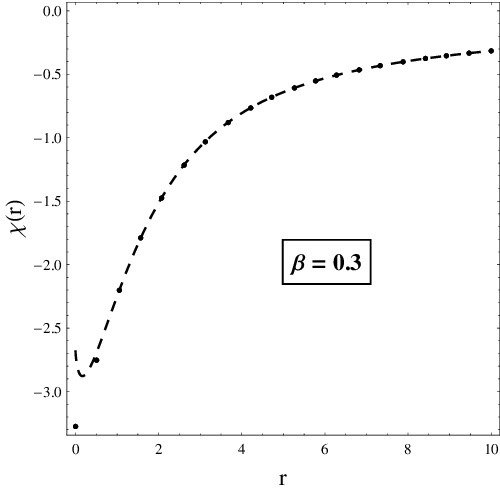}
\caption{\label{FigPert} \sl Plot of $ \chi (r)$ for different values of $\beta$ ($m=1,\, \mu=1.2$). 
The dots indicate the numerical evaluation of eq.~\eq{Chi-Int}, while the dashed line corresponds to the approximated result from the perturbation theory resulting from the sum of eqs.~\eq{Pot4d} and \eq{fina}.
}
\end{center}
\end{figure}

Another approach to obtain the potential with logarithmic quantum corrections is to expand \eq{Chi-Int} in the parameter $\beta$. This relies on the assumption that the scale related to the one-loop quantum correction term is, consistently with the perturbative loop expansion, much smaller than the classical counterpart because the former is of order $O(\hbar)$.  
Therefore, we define the perturbative expansion of $\chi$ in the parameter $\beta$ as follows,
\begin{equation} 
\n{eq2}
\chi = \chi^{(0)} + \chi^{(1)} + O(\beta^2),
\end{equation}
where $\chi^{(0)}$ is the classical potential, whose explicit solution is given by eq.~\eqref{Pot4d}, and
\begin{equation} 
\n{quan-po}
\chi^{(1)} (r) =  \frac{\be \ka M}{ \pi ^2 r} \int_0^\infty dk  \,\frac{k \sin (kr)}{(1 + \al k^2)^2}\,  \ln \left(  k/\mu \right) .
\end{equation}
is the first order quantum correction. 

To solve the integral in~\eq{quan-po} it is useful to apply the change of variables in the form $k = m x$, where $m=\sqrt{1/\al}$ (as in eq.~\eqref{Int4Der}).
Thus, we get:
\begin{eqnarray}
&& \chi^{(1)}(r)  = \frac{\be m^2 \ka M}{ \pi ^2 r} \, {I}^{(1)},
 \label{W_dpn}
\\
&& {I}^{(1)} \equiv \int_0^\infty dx  \,\frac{x \sin (mr x) \ln(mx/\mu)}{(x^2+1)^2}  .
\n{Q}
\end{eqnarray}
Because of the perturbative expansion, and differently from the approach of Sec.~\ref{Sec4}, here the logarithmic function appears in the integrand's
numerator, and it is possible to obtain the solution for the integral in~\eqref{W_dpn} using Cauchy's residue theorem.

Let us define
\beq
\n{eqt1}
F(z) = \frac{z \ln(m z/\mu)}{( z^2+1)^2} e^{imrz}, 
\quad - \frac{\pi}{2} \leqslant \,\text{arg}\, z < \frac{3\pi}{2}.
\eeq
The branch cut defined in \eq{eqt1} corresponds to the negative part of the imaginary axis. Therefore, one can define the oriented closed path $C$ depicted in~Fig.~\ref{con}, for which $\Im (z) \geqslant 0$. Notice that $C$ has an indentation around $z=0$, where $\ln z$ is not defined. Since only the double pole at $z=+ i\,$ is inside $C$, we find:
\beq
\n{eq3}
\ointctrclockwise_C dz\, F(z) = 2 \pi i \Res (F(z), i).
\eeq
On the other hand, using the paths described in Fig.~\ref{con} we get
\begin{equation} 
\begin{split}
\n{eq4}
& \hspace{-1.5 mm}
\ointctrclockwise_C dz \, F (z)  =  \int_\varepsilon^R dx \, \frac{x \ln(m x/\mu)}{( x^2+1)^2} e^{imrx}
+ \int_{C_R} dz \, F(z)
\\
& \hspace{1.0 mm}
+ \int_{-R}^{-\varepsilon} dx \, \frac{x [\ln (m|x|/\mu) + i \pi] }{(x^2+1)^2} e^{imrx}
+ \int_{C_\varepsilon} dz \, F(z)
.
\end{split}
\end{equation}
Taking $R$ sufficiently large and $\varepsilon$ small, we can use the
triangle inequalities and the upper bound for contour integrals over an arc $C_k$, 
\beq
\Bigg| \int_{C_k} dz \, F(z) \, \Bigg| \leqslant \mbox{length} \{C_k \} \times  \max\limits_{z \in C_k}  \, \{|F(z)|\}
,
\eeq
to find
\begin{equation}
\begin{split}
&
\Bigg| \int_{C_R} dz \, F(z) \, \Bigg|
\leqslant \frac{\pi R^2 \left[ \ln (mR/\mu) + \pi \right] }{(R^2-1)^2}
,
\\
&
\Bigg| \int_{C_\vp} dz \, F(z) \, \Bigg|
\leqslant \frac{\pi \vp^2 \left[\ln (m\vp/\mu) + \pi \right]}{(1-\vp^2)^2} .
\end{split}
\end{equation}
Therefore, the integrals along $C_R$ and $C_\vp$ vanish, respectively, in the limit $R \to \infty$ and $\vp \to 0$. 
Thus, making the substitution $x \mapsto - x$ in the third integral in the right-hand side of formula~\eq{eq4} 
and equaling with~\eq{eq3} we find
\begin{equation}
\begin{split}
&
\hspace{-0.1mm}
2 \pi i \Res (F(z), i) =
\int_{0}^{\infty} dx \,  \frac{x (e^{imrx} - e^{-imrx}) }{(x^2+1)^2}  \ln (mx/\mu)
\\
& \qquad \quad - i \pi \int_{0}^{\infty} dx \,  \frac{x e^{-imrx}}{(x^2+1)^2} 
.
\end{split}
\end{equation}
Now, having the definition~\eq{Q} in mind, after a small rearrangement, the imaginary part of the equation above gives
\beq
\n{K-an}
{I}^{(1)}
= \pi \Re [ \Res (F(z), i) ] +  \frac{\pi}{2} \int_0^\infty dx \, \frac{x \cos (mr x)}{(x^2+1)^2} 
.
\eeq
Like in the case discussed at the beginning of Sec.~\ref{Sec4}, the integration over the negative real axis leaves a remaining integral, but the one found here is more tractable than the other. Actually, the last integral in \eq{K-an} can be reduced, after integration by parts, to the Raabe's integral (for $m^2\geqslant0$)~\cite{book2},
\beq
\begin{split}
\n{Raabe}
\int_0^\infty dx \, \frac{2 x \cos (m r x)}{(x^2+1)^2} =& \, 1 + \frac{mr}{2} [ e^{mr} \text{Ei} (-mr) 
\\
&
- e^{-mr} \text{Ei}(mr) ]
,
\end{split}
\eeq
where $\text{Ei}(x)$ is the exponential integral function,
\begin{equation}
\text{Ei}(x) = - \int_{-x}^\infty dt \, \frac{e^{-t}}{t}.
\end{equation}
It is not difficult to show that
\beq \qquad
\Re [\Res (F(z),i)] = \frac{e^{-m r}}{4}   [ m r \ln  ( m/\mu ) - 1 ] 
.
\eeq

Therefore, collecting all the results, we get the solution for the quantum correction to the classical potential,
\begin{equation}
\begin{split}
\n{fina}
\chi^{(1)}(r) =& \,\,  \frac{\be \ka M m^2}{4 \pi r} \Bigg\{ 
1  - [ 1 - m r \ln (m/\mu)] e^{- m r} 
\\
& + \frac{m r}{2} \left[  e^{m r} \, \text{Ei}(-m r) - e^{- m r} \, \text{Ei}(m r)  \right] 
\Bigg\}
.
\end{split}
\end{equation}

In Fig.~\ref{FigPert} we present the comparison of the numerical solution of~\eq{Chi-Int} and the perturbative one-loop approximation~\eq{fina}. It reveals that, close to $r=0$, the $O(\hbar)$-result deviates from the non-perturbative one as $\be$ increases. Not surprisingly, this represents the one-loop approximation breakdown for large $\be$ in the high-energy domain. We remind that, within the perturbation theory, even the numerical solution of eq.~\eq{Chi-Int} is not accurate beyond $O(\hbar)$, for the form factor in the action~\eq{act} does not include the corresponding higher-loop terms. 

Finally, as mentioned in the previous section, we can use~\eq{fina} to investigate the large-$r$ limit of the quantum correction to the potential.
Since the term inside the square brackets in the last line of eq.~\eqref{fina} behaves as
$$
e^{m r}  \text{Ei}(-m r) - e^{- m r}  \text{Ei}(m r) =  -\frac{2}{m r}-\frac{4}{(m r)^3} -\frac{48}{(m r)^5}+ O\big( r^{-7}\big),
$$
it is straightforward to verify that 
\begin{equation} \label{limitIR}
\chi^{(1)}(r) \underset{r \to \infty}{\sim } - \frac{\be \ka M}{ 2 \pi r^3} .
\end{equation}
This shows that in the far-IR limit, the one-loop corrected potential matches the behavior of the quantum corrected potential evaluated in~\cite{Duff:1974ud,Donoghue:1993&94,Muzinich:1995uj,Hamber:1995cq,Dalvit:1997yc,Akhundov:1996jd,Khriplovich:2002bt,BjerrumBohr:2002kt} in the realm of the effective quantum theory of general relativity. Our result supports the hypothesis of the universality of the IR quantum gravity approach. 


\section{Conclusion}
We computed the quantum corrections to the Newtonian potential and all the other  gravitational perturbation components in the quantum effective action for the fourth-order gravitational theory. The calculations were done both numerically and perturbatively, but also analytically close to the singularity point $r = 0$.

We proved that the logarithmic quantum corrections improve the behavior of the Kretschmann curvature invariant near $r=0$ respect to the classical theory, but are not enough to solve the spacetime singularity problem.  

\acknowledgments
This work was supported by the Basic Research Program of the Science, Technology and Innovation Commission of Shenzhen Municipality (grant no. JCYJ20180302174206969).



\end{document}